\documentclass[conference]{IEEEtran}
\usepackage{blindtext, graphicx}

\usepackage{cite}
\usepackage{graphicx}
\usepackage[cmex10]{amsmath}
\usepackage{subcaption}
\usepackage{dblfloatfix}
\usepackage{balance}
\usepackage{url}

\begin{document}
%
\title{Learning Seasonal Phytoplankton Communities with Topic Models}

\author{\IEEEauthorblockN{Arnold Kalmbach\IEEEauthorrefmark{1},
        Heidi M. Sosik\IEEEauthorrefmark{2},
        Gregory Dudek\IEEEauthorrefmark{1}, and,
        Yogesh Girdhar\IEEEauthorrefmark{3}}%
 
        \IEEEauthorblockA{\IEEEauthorrefmark{1}Centre for Intelligent Machines, School of Computer Science,\\
        McGill University, Montr\'{e}al, QC, Canada, {\tt\small \{akalmbach,dudek\}@cim.mcgill.ca}
        }%
        \IEEEauthorblockA{\IEEEauthorrefmark{2}Biology Department, Woods Hole Oceanographic Institution, \\ Woods Hole, MA, {\tt\small hsosik@whoi.edu}
        }%
        \IEEEauthorblockA{\IEEEauthorrefmark{3}Applied Ocean Physics and Engineering Department, Woods Hole Oceanographic Institution\\
        Woods Hole, MA, {\tt\small yogi@whoi.edu}
        }
}



\maketitle



\begin{abstract}
In this work we develop and demonstrate a probabilistic generative model for phytoplankton communities. The proposed model takes counts of a set of phytoplankton taxa in a timeseries as its training data, and models communities by learning sparse co-occurrence structure between the taxa. Our model is probabilistic, where communities are represented by probability distributions over the species, and each time-step is represented by a probability distribution over the communities. The proposed approach uses a non-parametric, spatiotemporal topic model to encourage the communities to form an interpretable representation of the data, without making strong assumptions about the communities. We demonstrate the quality and interpretability of our method by its ability to improve performance of a simplistic regression model. We show that simple linear regression is sufficient to predict the community distribution learned by our method, and therefore the taxon distributions, from a set of naively chosen environment variables. In contrast, a similar regression model is insufficient to predict the taxon distributions directly or through PCA with the same level of accuracy.
\end{abstract}

\IEEEpeerreviewmaketitle

\section{Introduction}

\begin{figure}[t]
\includegraphics[width=0.9\columnwidth]{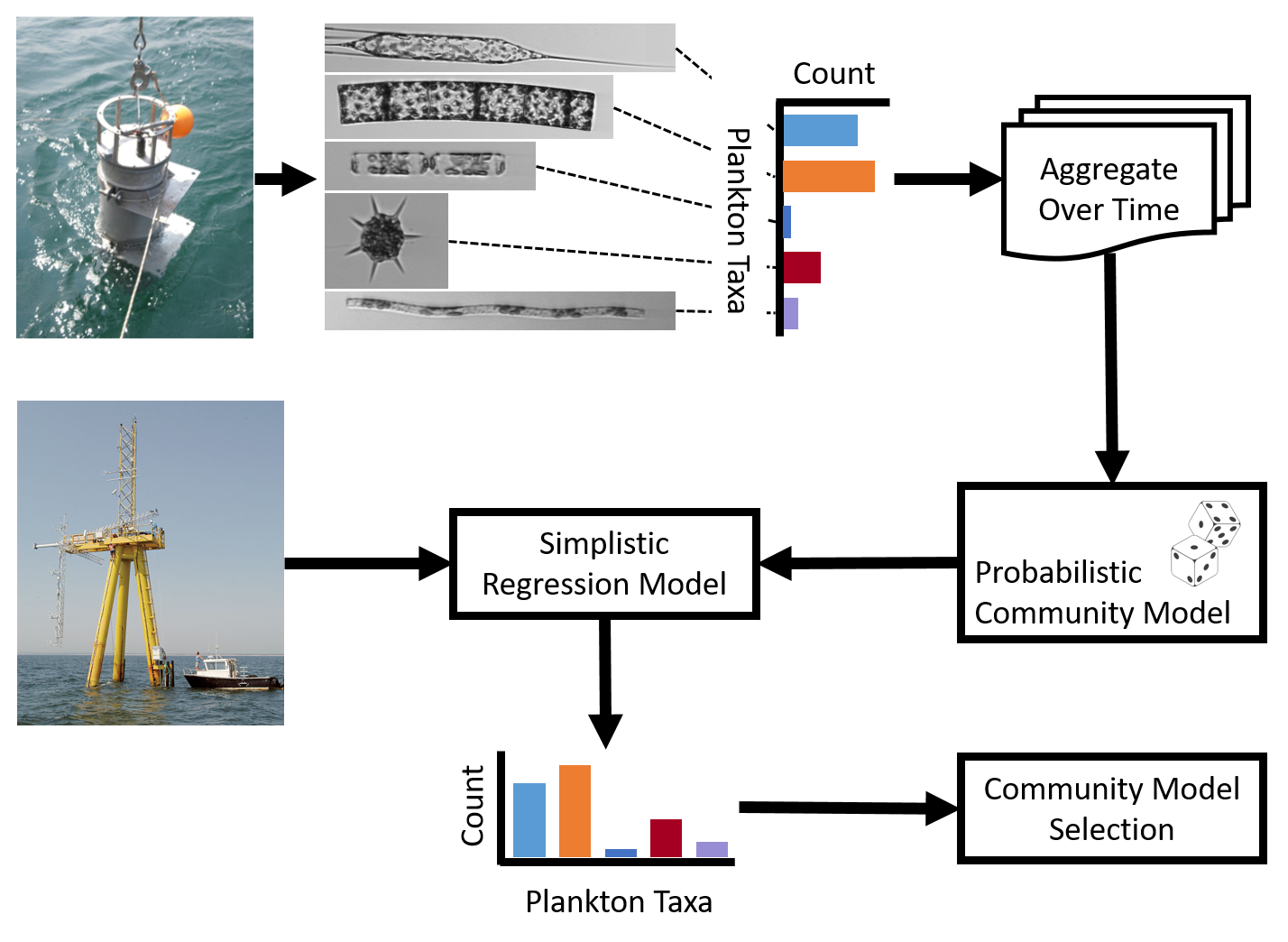}
\caption{\emph{System Overview:} Imaging FlowCyto Bot (top left) automatically detects and classifies phytoplankton. We aggregate the counts of all the taxa on each day in long-term deployment of IFCB, and use this dataset to learn a probabilistic community model. We use adjacent sensors at Martha's Vineyard Coastal Observatory (12m Sea Node, shown bottom left) and fit a simple regression model, predicting taxon distributions from environment variables using the learned communities. By demonstrating that our method performs this task well, we show that our community model is both accurate and interpretable.}
\label{fig:overview}
\end{figure}


Phytoplankton are microscopic organisms that form the base of marine food webs.
They produce chlorophyll and other pigments to harvest sunlight and fuel photosynthesis, so they can utilize $\mathrm{CO}_2$ and other nutrients to produce $\mathrm{O}_2$ and new organic matter. As such, they play critical roles in global biogeochemical cycles and in structuring marine ecosystems.  
Marine scientists have long used techniques to measure the amount of chlorophyll in a water sample as a proxy for phytoplankton biomass \cite{Lorenzen1966}. These methods are coarse and give only bulk indices, with no information about which species of phytoplankton are present. Phytoplankton are extremely diverse, however, and their community structure plays a major role in shaping ecosystems and their functions. As an extreme example, particular species are known to cause toxic blooms that can threaten wildlife as well as human health.

To meet the gap in observational capability that includes taxonomic resolution, Sosik and Olson have developed the automated, submersible Imaging FlowCytobot (IFCB)~\cite{Olson2007a} and a coupled analysis system \cite{Sosik2007,Sosik2016}. This system can detect and classify phytoplankton automatically in small samples of ocean water collected serially over long periods of time (weeks to years). The taxa classified in the Martha's Vineyard Coastal Observatory (MVCO) deployment of IFCB, used in this work, show a remarkable level of detail, and are presented in Fig.~\ref{fig:stripplots},~(bottom).

In this work we present a model which factors the daily taxon count aggregates produced by the IFCB into a small number of communities. A key advantage of our method is that the communities it produces are sparse, and the daily community distributions it produces are temporally smooth. The result is that the learned representation of the data is more interpretable than standard decomposition methods such as PCA. It is our goal that community distributions should not only fit the data, but also provide a meaningful decomposition enabling further modelling and intuitive understanding. To this end we choose a simplistic regression model, trained to predict the community distributions, and therefore taxon distributions, from a basic set of oceanographic and meteorological variables. Rather than developing the perfect regression model or choice of variables, we choose a community model that allows for a very naive regression model to perform well. This ensures that the community model affords a simple yet accurate interpretation.

We demonstrate that our method achieves this goal by training our community model on an extensive dataset from a fixed IFCB deployment at the Martha's Vineyard Coastal Observatory (MVCO), in Cape Cod, Massachusetts. There are many choices of probabilistic community model which fit the data well, therefore we select among them using their performance on the regression task. Remarkably, the best community model shows strong seasonal patterns, despite the fact that our community model makes no such assumptions. We find that our approach is relatively accurate in predicting the original taxon distributions from environment data. In contrast, standard approaches such as direct regression, or regression on a PCA decomposition of the taxon distributions do not achieve the same level of accuracy with the basic regression model.

Our recent work \cite{Kalmbach2017} proposed the use of a similar phytoplankton community model for robust detection of hotspots of a spatially distributed plankton taxa. This work instead focuses on modeling community distribution of plankton observed at a stationary location, over a long period of time, and using the model to predict plankton distribution based on environmental conditions. 


\section{Approach}

\begin{figure}
\includegraphics[width=\columnwidth]{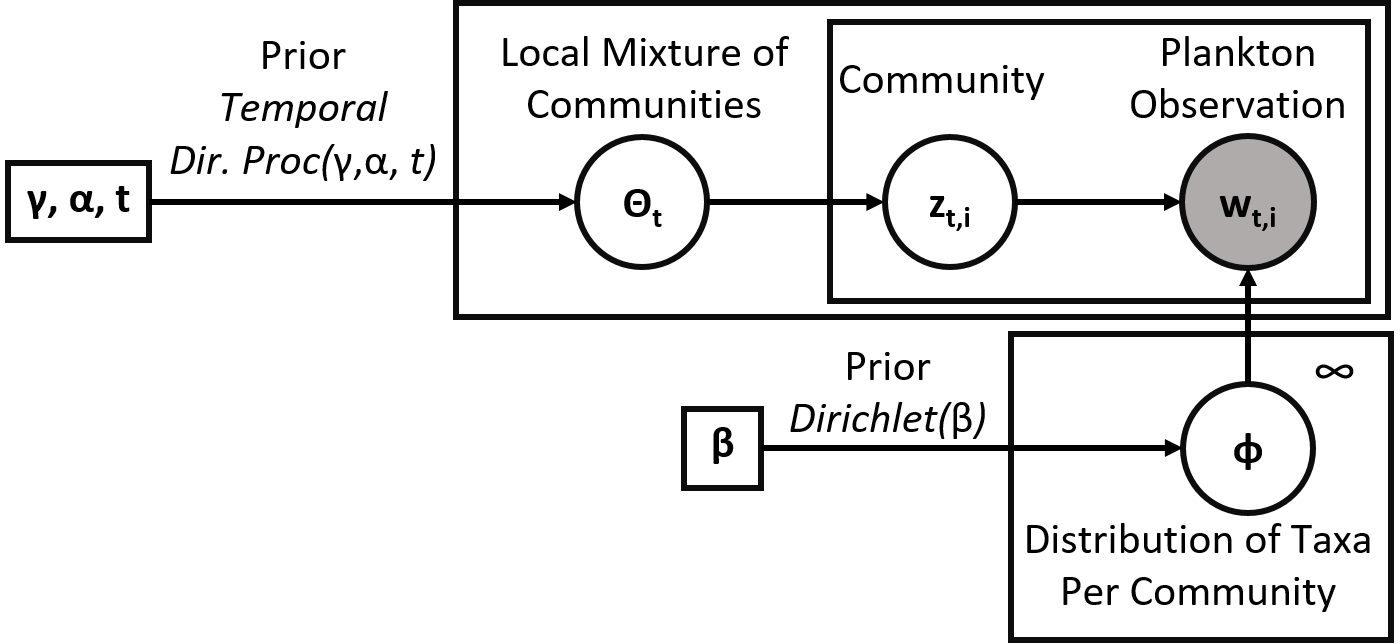}
\caption{The graphical model used by our community model.}
\label{fig:plate}
\end{figure}

The proposed phytoplankton community model is adapted from a Bayesian non-parametetric topic modelling approach to discover common co-occurrence patterns in the taxa count data. The observed taxon distribution at each time step is modelled by a mixture of communities (topics), and each community is a probability distribution of the taxa. Successful topic modelling approaches from the text modelling literature often use Dirichlet priors to encourage topics to be sparse. Analogously, we use Dirichlet priors in our model to ensure that each community has a small number of taxa with positive probability \cite{Blei2012}. For the mixture of communities at each time step, we use a (spatio) temporal Dirichlet process prior \cite{Girdhar2015a}. This prior encourages the community mixtures to be sparse, similar to a standard Dirichlet prior, but also causes the community mixtures to be temporally smooth, and avoids the need know the number of communities \emph{a priori}.

The full model structure is shown in Fig.~\ref{fig:plate}. Learning in this model involves finding a latent community assignment $z_{i,t}$ for each individual plankton observation $w_{i,t}$. The maximum likelihood parameters $\Theta, \Phi$ are estimated jointly, such that 
$P(z_{i,t} = k | t) = \Theta_{t,k}$, and  $P(w_{i,t} = v | z_{i,t} = k) = \Phi_{k,v}$. After learning, the maximum likelihood taxon distributions under the model can be trivially recovered by the product $\Theta \Phi$.

Rather than evaluating the quality of different community models by the accuracy of their maximum likelihood taxon distributions, which can be made arbitrarily good on the training set by introducing more communities, we train a simplistic regression model on the community distributions. We take a set of environment variables, averaged over the same daily observation windows used for plankton count aggregation. We reject outliers based on the median absolute deviation of each variable. Then we standardize each variable independently, subtracting its average and dividing by its standard deviation. Finally, we fit a linear ridge model of the community distributions from the standardized environment variables.

After training, we can take a set of environment measurements, shift and scale them using the standardization parameters, predict the community distribution using the learned weights, and finally predict the maximum likelihood taxon distributions given the predicted communities. If the maximum likelihood taxon distribution for a given set of communities is not accurate, then this regressor will perform poorly, reflecting the fact that the community model did not work well. However the regression model will also perform poorly if the community model is accurate, but the presence of certain communities is too complicated to predict by a linear function of simple environement variables. In contrast, the best performing community models on the regression task are both accurate and simple enough to be interpreted naturally.

\section{Experiment}

\begin{figure*}[]
    \centering
    \begin{subfigure}[t]{\textwidth}
        \centering
        \includegraphics[width=\columnwidth]{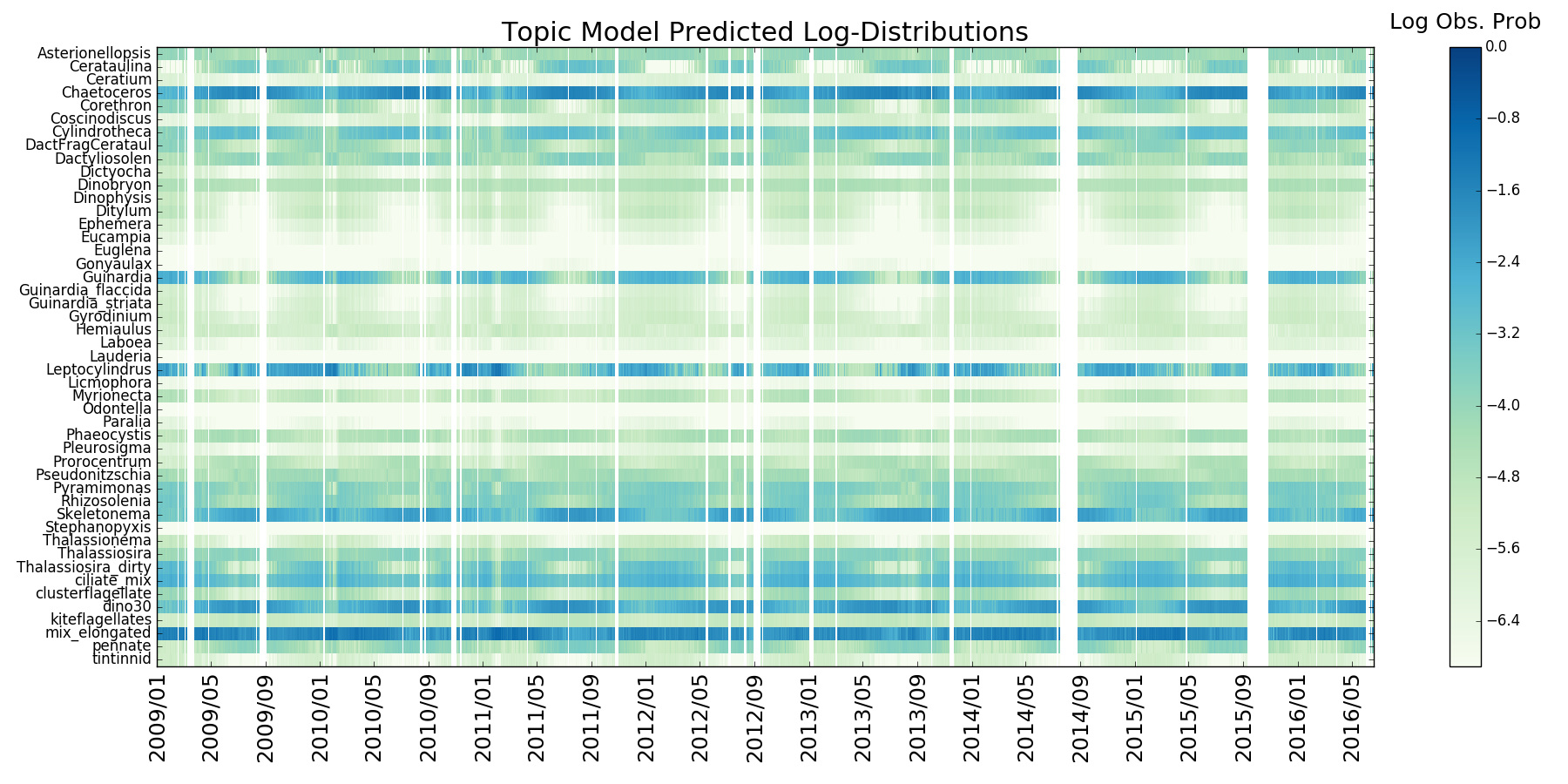}
    \end{subfigure}%
    \\ 
    \begin{subfigure}[t]{\textwidth}
        \centering
        \includegraphics[width=\columnwidth]{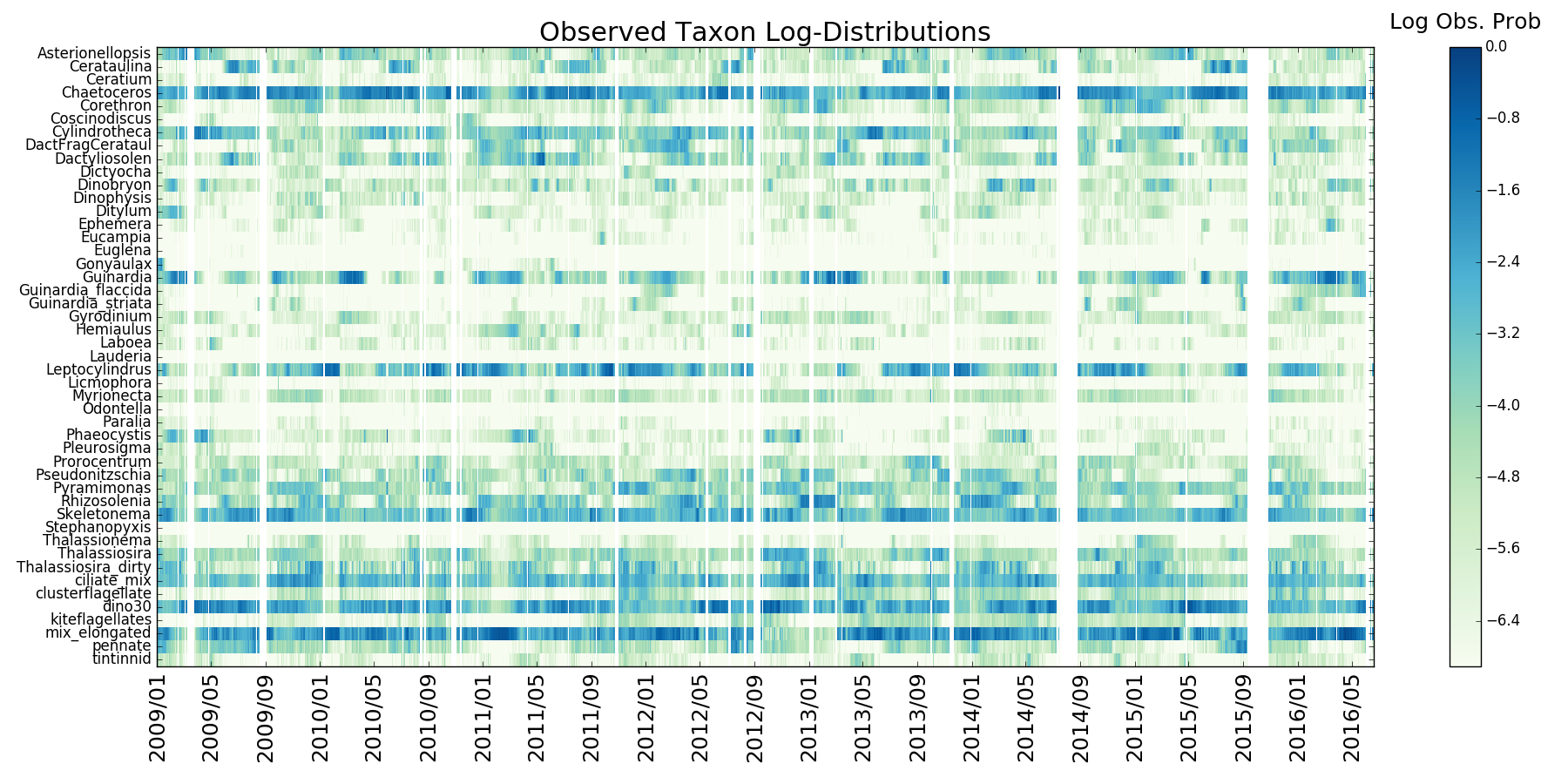}
    \end{subfigure}
    \caption{ The time series plot shows the distribution of predicted plankton taxa (top) and observed groundtruth taxa distribution (bottom). We see that the predicted distribution is able to capture most of the low frequency structure of the observed data. The ground-truth data is publicly available at \protect\url{http://ifcb-data.whoi.edu/mvco/}}
    \label{fig:stripplots}
\end{figure*}

We demonstrate our method on a dataset recorded continuously from Jan. 2009 to Jul. 2016 at the MVCO. The IFCB was configured to automatically sample from 5 ml of surface seawater approximately every 20 min. The classification system generated an average of over 1100 observations per day, distributed over 47 taxa (Fig.~\ref{fig:stripplots},~bottom).

\begin{figure}[h!]
\includegraphics[width=\columnwidth]{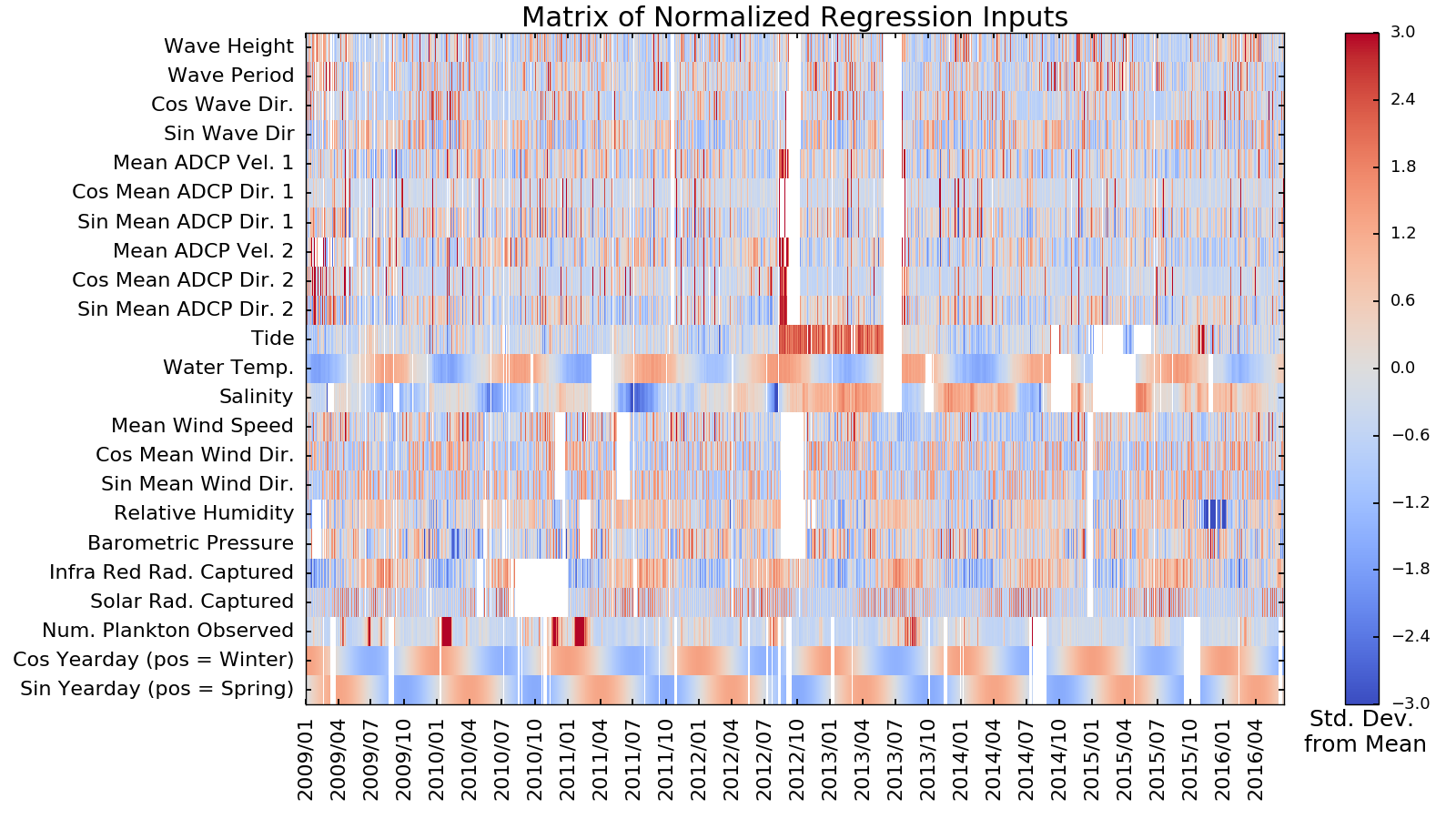}
\caption{Inputs to our regression system, oceanographic and meteorological Variables recorded at the Martha's Vineyard Coastal Observatory over the period Jan. 2009 to Jul. 2016. Presented as our system receives them, centered and scaled to a normal distribution. White spaces indicate gaps in the data or where outlier data was removed.
This data is publicly available at \protect\url{http://www.whoi.edu/mvco/data}}
\label{fig:regression_inputs}
\end{figure}

We aggregated the observations to produce the taxon distribution for each day during the 7.5 year period, and used this as input to our method. For the regression model, we chose a suite of 18 environment variables from the MVCO ocean data and meteorological data summaries (Fig.~\ref{fig:regression_inputs}). In addition to being naively chosen, the environment data features significant gaps and systematic noise due to the practical challenges of long-term ocean sensor deployments. Much of the systematic noise was suppressed by our outlier rejection method, yet the regression task remains extremely challenging.

We trained our community model on the taxon distributions over the entire period multiple times for different hyperparameter settings. The hyperparameter search process involved varying $\alpha$ (prior for sparsity of community distribution on each day), $\beta$ (prior for sparsity of taxa in each community), $\gamma$ (prior for data complexity), and $g$ (prior for temporal smoothness of community distributions). For each community model corresponding to a different combination of choices of these hyperparameters, we trained the regression model 8 times, once using each year as the test set and the other 6.5 years for training. Within each training set, we chose the regularization parameter for ridge regression using hold-one-out cross validation. Finally, we used the resulting regressors to predict the respective held-out community distributions for each year, and used the respective community models to predict the taxon distributions.

We demonstrate the utility of our method in comparison to two more standard regression techniques. The first is to predict the taxon distributions directly with a similar ridge regression model and training procedure. The second is to first take a PCA decomposition of the taxon count data, using the first $K$ principle components, where $K$ is the same as the number of communities used by our model, and then use the same ridge regression model and training procedure to predict the PCA weights, and finally project the weights back to predict the taxon distributions.

\section{Results}

\begin{figure}
\centering
\includegraphics[width=\columnwidth]{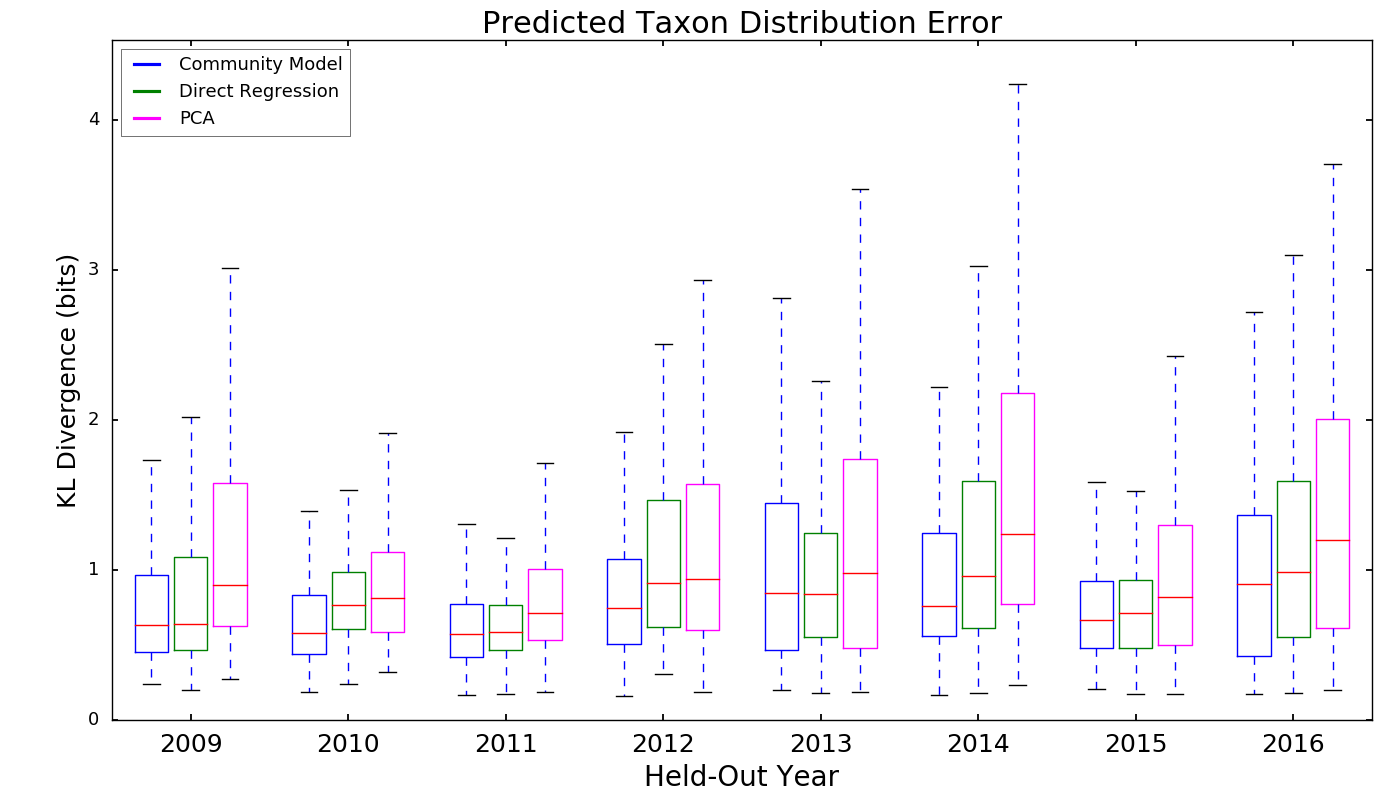}
\caption{Comparison of daily taxon distribution prediction errors, for each of the three regression methods, for each of the years in the dataset. For each year, the other 6.5 years were used as training data. Note that for each year the Community model based regression method (ours, left) shows the lowest median KL-Divergence (lower = more accurate).}
\label{fig:regression}
\end{figure}

We evaluated the resulting regression systems for all hyperparameter settings as well as the two baseline methods by comparing the predicted taxon distributions to the true distributions on each day in the dataset. Our error measure is the KL-Divergence between the predictions and the held-out distributons\footnote{KL-Divergence is a measure of the information lost when approximating one distribution with another, equal 0 if the distributions are identical and $\infty$ if they do not have the same support. If we sample a taxon randomly from the estimated distribution, and that taxon has probabilities $p,~q$ under the estimated and true distributions respectively, the KL-Divergence gives the expected value of $log \frac{p}{q}$.}. We chose the community model with the lowest average KL-Divergence over all the days in the dataset, ultimately picking a model with 6 active communities. The predicted taxon distributions for this model are shown in Fig.~\ref{fig:stripplots}~(top).

Fig.~\ref{fig:regression} shows the taxon distribution prediction errors for this community model and our two baseline models, broken out by year. The boxes represent the distribution of prediction errors for nearly 365 days in 2009 through 2015, and 172 days in 2016 (nearly 365 because of some small gaps in the taxon count data). Our community model (leftmost for each year) achieved the lowest median error on every year in the dataset. We found that by optimizing the hyperparameters for the regression task, we were able to choose an interpretable representation of the community structure. In contrast, PCA does not feature any prior for temporal smoothness. As a result although its prediction error is on average only a little less accurate than our model's, the sequences of predictions it makes are sometimes implausible, featuring taxon distributions which change much more rapidly than the observed data. We found that both baselines were extremely susceptible to noise in the environment data, on average performing better than expected, but occasionally making extremely poor predictions (Fig.~\ref{fig:example_timeseries}). With our model, the regression problem is of a lower dimensionality than for direct regression, and therefore less susceptible to overfitting. For this reason when both models are presented with the same small amount of training data, our model is more able to avoid large errors for new inputs unlike the training data.

\begin{figure}
    \centering
    \begin{subfigure}[t]{\columnwidth}
        \centering
        \includegraphics[width=\textwidth]{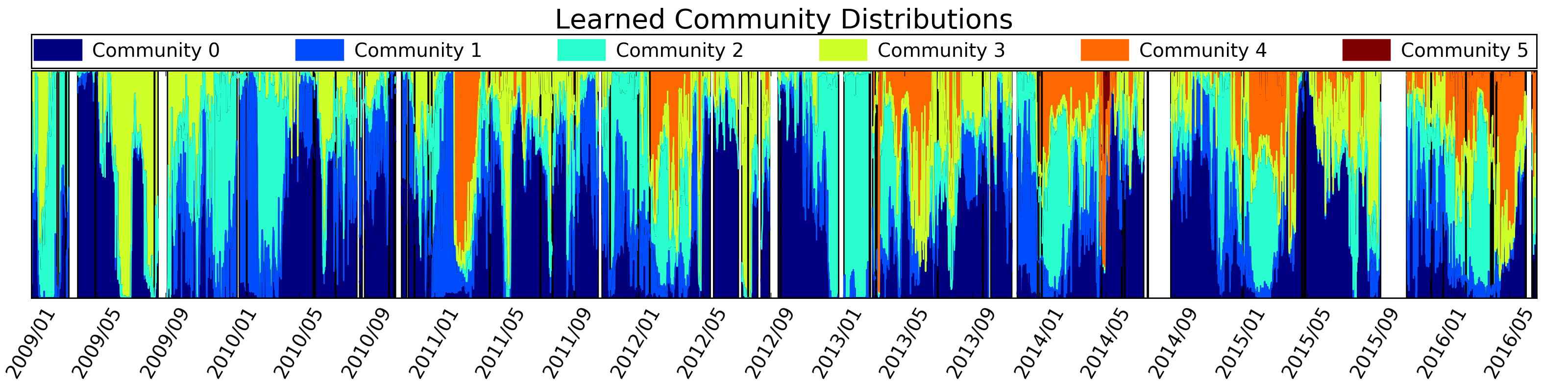}
        \caption{Daily community distributions over 7.5 years for the best performing community model on the regression task.}
        \label{fig:theta_stacked}
    \end{subfigure}%
    \\ 
    \begin{subfigure}[t]{\columnwidth}
        \centering
        \includegraphics[width=\textwidth]{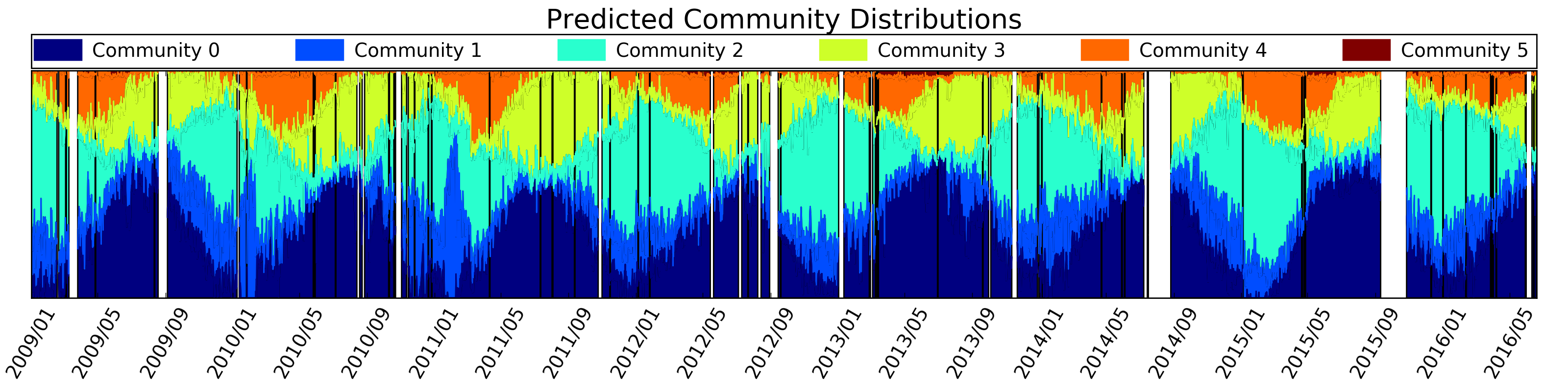}
        \caption{Daily community distributions predicted from environment data. For each year, the model was trained to predict the community distributions based on the other 6.5 years of environment and community data.}
        \label{fig:theta_hat_stacked}
    \end{subfigure}
    \\
    \begin{subfigure}[t]{\columnwidth}
        \centering
        \includegraphics[width=\columnwidth]{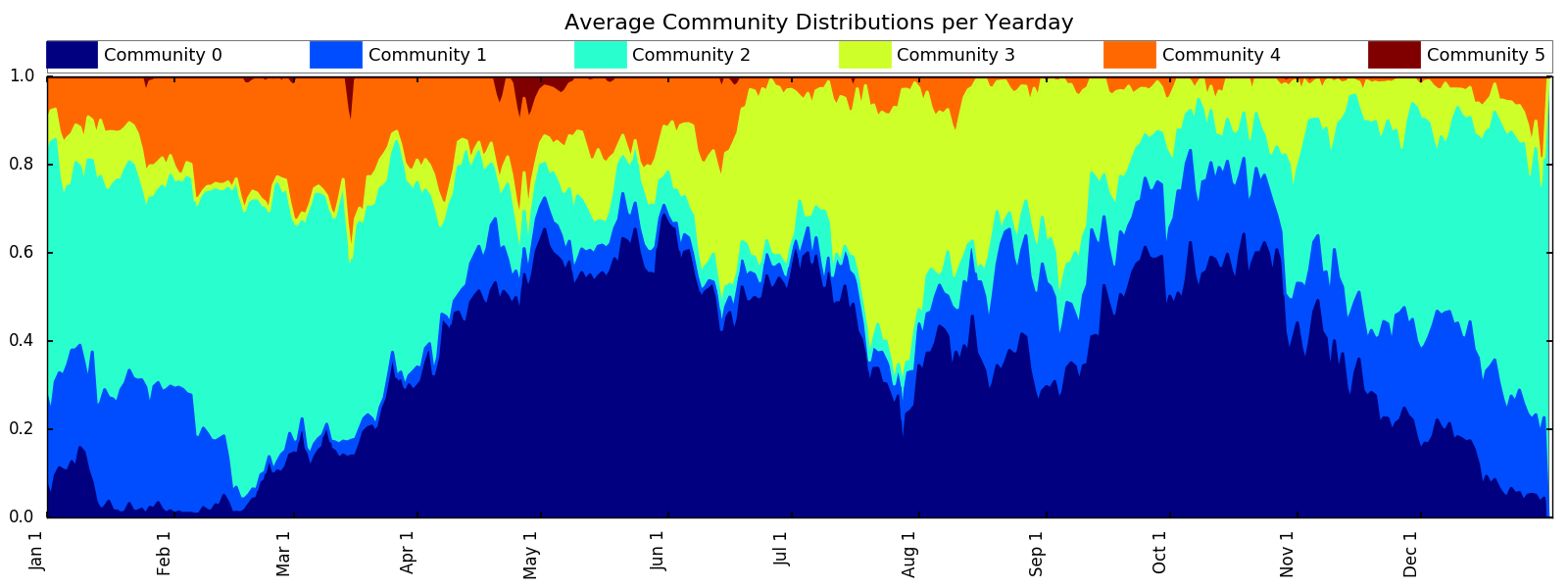}
        \caption{Average community distribution for each day of the year over the entire dataset.}
        \label{fig:theta_mod}
    \end{subfigure}

    \caption{The learned and predicted community distributions. The horizontal axis represents time, and each community is represented by a color. The fraction of observations on a day belonging to a particular community are shown by the size of the colored area (totalling 1 for each day). Our model finds strong seasonal structure in the data without having such an assumption built-in. }
    \label{fig:theta}
\end{figure}

An intriguing result of our community regression model is that nearly all of the magnitude in the weights of the learned regression parameters is either on the day of the year, the water temperature, or the number of plankton classified for a given day (Fig.~\ref{fig:weights}). Note that we have used a $cos$,~$sin$ pair to encode cyclic variables. In the case of the day of the year, the positive cosine direction encodes Winter and the negative Summer, while the positive sine encodes Spring and the negative encodes Fall. From our regression matrix, we can see the interpretations `community 0 will be found in the Summer and early Fall, especially when the water temperature is lower than usual`, `community 1 will be found in the Winter, especially when the number of plankton is high', `community 2 will be found mostly in the Winter and Spring, when the total number of plankton is lower', `community 3 will be found mostly in the Summer and Fall, but with lower probability than community 0', and `community 4 will be found in the late Winter and Spring, when the water temperature is slightly higher than usual'. We found that the best performing community models showed strong seasonal structure, rather than relying on other variables. This is made evident by the average community distribution for each day of the year over the entire dataset, before regression (Fig.~\ref{fig:theta_mod}). Note that there are many possible community decompositions, and although our model makes weak assumptions about the temporal smoothness of the communities, it does not have any prior knowledge of the seasonal nature of the data. By performing hyperparameter optimization over the downstream regression task, we were able to select a model with just the right level of sparsity and temporal smoothness to emphasize this seasonal aspect and describe the data in an interpretable way.


\section{Discussion}

\begin{figure}[h!]
    \centering
    \begin{subfigure}[t]{0.5\textwidth}
        \centering
        \includegraphics[width=\columnwidth]{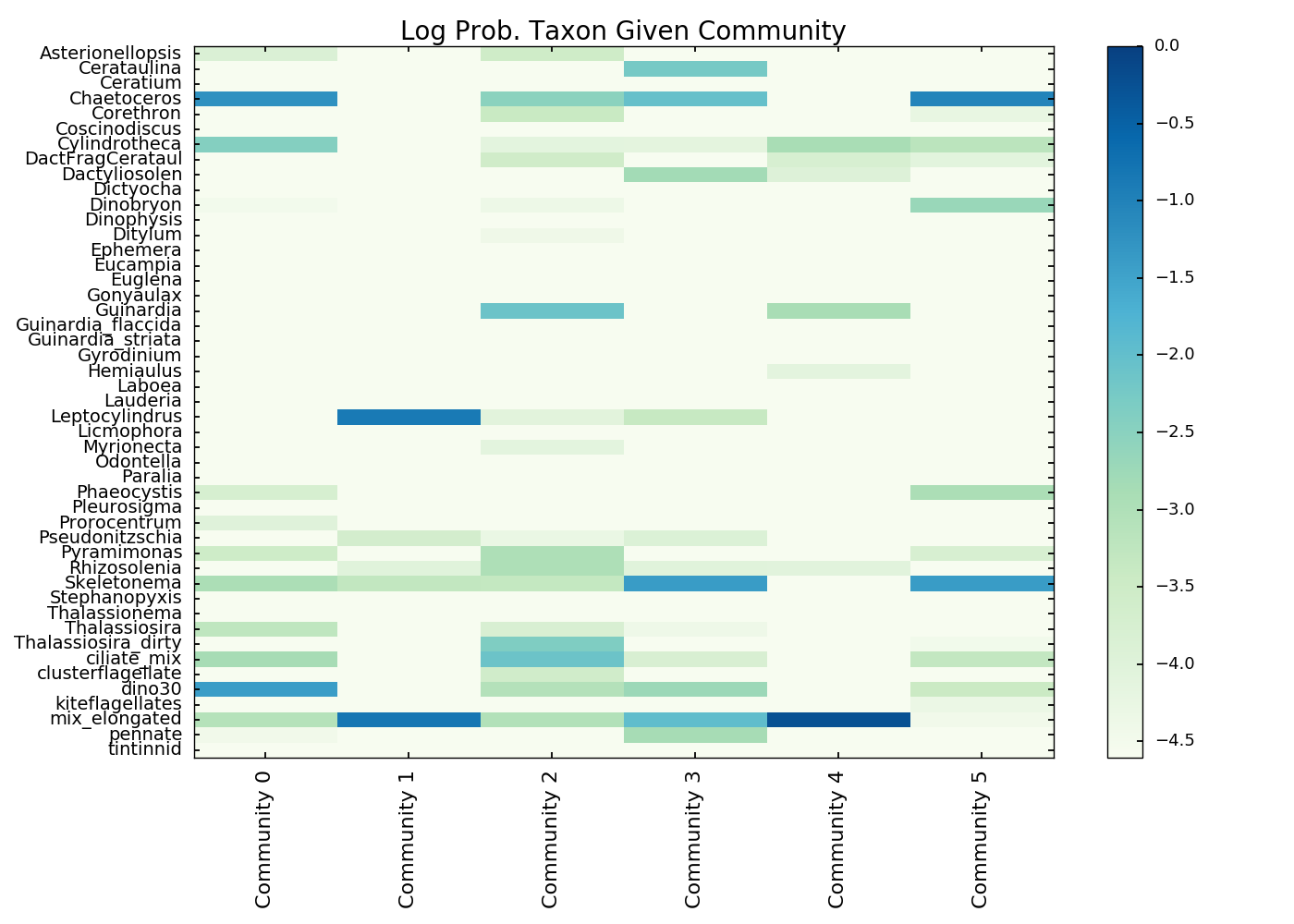}
        \caption{Each community represents a set of plankton taxa which are found to covary. Our model uses a Dirichlet prior to ensure that only a few taxa are active in each community.}
        \label{fig:phi}
    \end{subfigure}%
    \\
    \begin{subfigure}[t]{0.5\textwidth}
        \centering
        \includegraphics[width=\columnwidth]{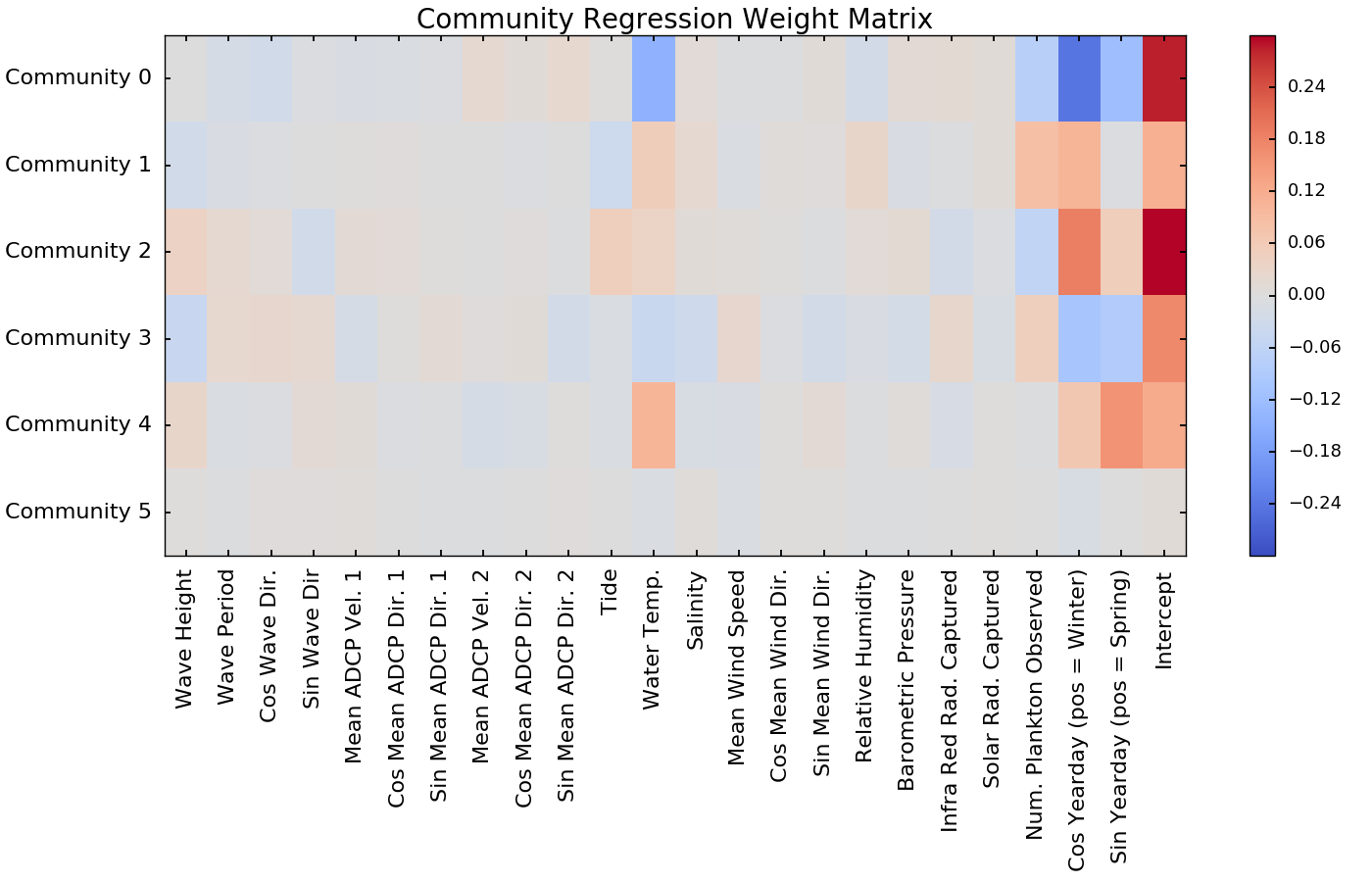}
        \caption{The weights used by our linear regression model to predict communities from environment variables. Note that nearly all of the magnitude is on a small subset of the variables, meaning that we can explain each community very simply.}
        \label{fig:weights}
    \end{subfigure}
    \caption{ Our model is not only an interesting demonstration of a general method, it also produces valuable data for understanding the factors involved in Phytoplankton ecology.}
    \label{fig:model_outcomes}
\end{figure}

All three of the regression models, and especially the community based model are able to make remarkably good predictions of the taxon distributions using mainly the day of the year. This indicates that the taxa observation count data is strongly seasonally structured. Nevertheless, the regression models lose a significant amount of information, even in the case of the reduced dimensionality community prediction problem (Fig.~\ref{fig:theta_hat_stacked}). In particular, the regressor captures most of the low-frequency variation in the communities because it can be explained by the season, but much of the high-frequency variation is lost, and as a result, the predicted community distributions are less sparse than the true community distributions.

An interesting area of future work is in performing a similar experiment with a more realistic regression model. Clearly, the performance of the ultimate regression models could be improved by careful choice of more factors and higher order terms which are known to be related to phytoplankton productivity. It is likely that overall performance could be increased by using more complex models as well, for instance a feedforward artificial neural network, or other popular machine learning techniques. In particular, sequential models, which predict tomorrow's community distribution from today's community distribution and a set of environmental factors are a promising method to incorporate population dynamics into our system. Despite the myriad opportunities for improving this portion of the model, the goal of regression in this experiment is to ensure that there is a simple interpretation of the communities, not only to accurately predict the data. There is a balance to be struck between the complexity of the model and how easy to interpret it is. In addition, more complex models run a higher risk of overfitting and require more data to train accurately, limiting their practical applications.

\balance

A further outcome of our experiment is the communities themselves learned by our model (Fig.~\ref{fig:phi}). We found that across different hyperparameter choices, the top few most active communities were relatively similar to those presented here. Some associations based on our model have ready explanations. For instance community 1 is dominated by the taxons ``mix\_elongated'', representing miscellaneous centric diatiom chains, and ``leptocylindrus'', which both exhibit elongated morphologies and easily confuse the IFCB's vision-based classification system. As a more exciting example, communities 2 and 4 are the only communities with significant probability of observing the taxon ``Guinardia'', and are predicted by warm water temperatures, while \emph{Guinardia delicatula} populations have been noted to be negatively associated with parasites that do not survive during cold winters \cite{peacock2014parasitic}. In our future work we hope to find more such examples which support the validity of our community model, and use our model to clarify less well-understood aspects of phytoplankton population ecology.

\begin{figure*}
    \centering
    \includegraphics[width=0.95\textwidth]{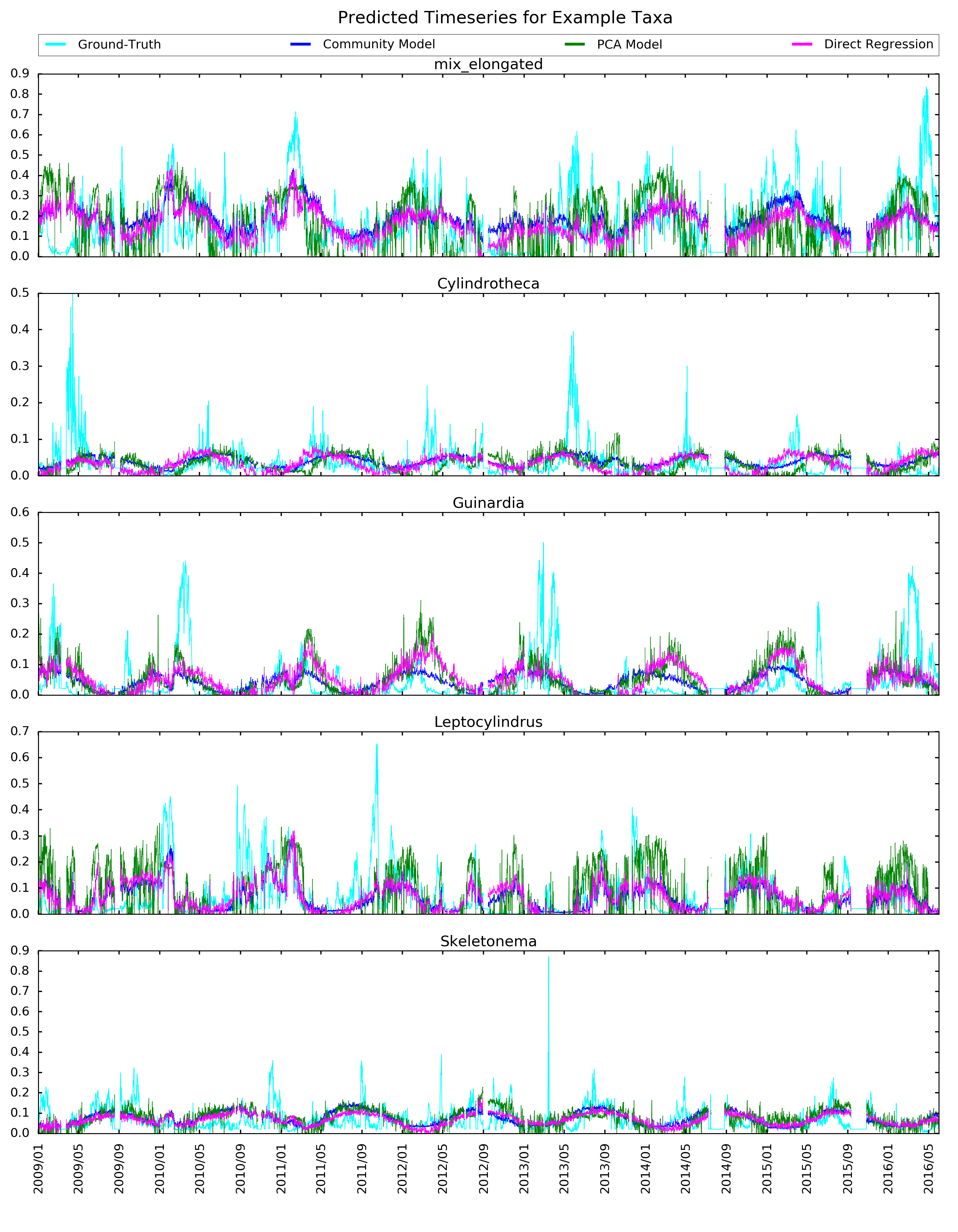}
    \caption{Probability a random observation on a given day was a particular taxon, for 5 examples of highly abundant taxa. The observed data, and regression outputs for each of the three methods are shown. Note that the predictions of community model regression are relatively smooth, while PCA and Direct regression vary
    unrealistically rapidly.}
    \label{fig:example_timeseries}
\end{figure*}

Finally, our community model takes a complex, high-dimensional population dataset, and offers immediate questions to pursue related to the ecology of specific taxa near MVCO. For instance, in Fig.~\ref{fig:theta_stacked} we see that community 4 begins to appear a little earlier each year from 2011, at first appearing only in the spring, and by 2015 persisting throughout the Winter. If we were to only look at the timeseries for the individual taxa, ``mix\_elongated'', ``Guinardia'', and ``Cylindrotheca'' which dominate community 4, this pattern is not obvious at all (See Ground-Truth, Fig.~\ref{fig:example_timeseries}). However, our model highlights that these taxa often co-occur, and that the pattern of co-occurrence shifts over the years. We hope to stimulate discussion and explanation of this and similar phenomena found in this dataset.

\section{Conclusion}

In this work we have presented a novel probabilistic generative model for phytoplankton communities. The model and hyperparameter selection procedures are designed with the goal of being simultaneously accurate, in terms of losing minimal information about the true distribution of phytoplankton species in a given time window, and interpretable, in terms of the presence of a distribution of communities being simple to predict from a set of naively chosen environment variables. We demonstrated our model on an extensive public dataset featuring counts of 47 plankton taxa and 18 environment variables recorded over 7.5 years at MVCO. We found that our method produced more interpretable representations of the count data than PCA or direct regression. Our model provides a novel general way to understand high-dimensional datasets of discrete taxon observations, as well as intriguing observations about the populations of phytoplankton near MVCO.

\section*{Acknowledgement}
\small{This work was supported in part by awards to YG from NOAA through its Cooperative Institute for the North Atlantic Region (CINAR) program, and  from WHOI; and to HMS from NASA's Ocean Biology and Biogeochemistry Program, and from NOAA through CINAR. We are indebted to Emily Brownlee for expert assistance with IFCB data collection and Joe Futrelle for facilitating IFCB data access and analysis workflows. We also thank the captain and crew of the Research Vessel Pisces and scientists from NOAA's Northeast Fisheries Science Center for enabling our participation in EcoMon surveys. We gratefully acknowledge the support via grant to GD of the Natural Sciences and Engineering Research Council of Canada (NSERC).}

\bibliographystyle{IEEEtran}
\bibliography{girdhar,arnold}

\end{document}